# The propagation of the light beam at the interface between uniform media and periodic media

Dmitry Levko

**Abstract:** We considered the wave propagation between two medias. For description of physical model choose nonlinear Schrödinger equation with saturation parameter. The solutions of respectively equation are found and analyzed.

## Introduction

At the present time nonlinear Schrödinger equation (NSE) find it's applications in different physical contexts: in nonlinear optics [1]-[3], in plasma's physics [4], in fiber optics [5], in Bose-Einstein condensates [6]-[7] and others. This equation admits the soliton solutions.

The key role solitons play in recent investigations in the study of wave propagation in photonic crystals [5]. For physical description of processes in such medium different authors take different versions of NSE [3], [8]-[9].

In [3] Yu. Kartashov and authors discovered the existence of surface gap solitons supported by the surface between optical lattice and uniform media imprinted in Kerr-type nonlinear media.

In [9] the surface defect gap solitons were discovered. These waves are the nonlinear defect modes that bifurcate out from the linear defect modes. It has the potential applications for routing of optical signals.

In periodic lattices solitons form when their propagation constant lies within certain regions [10] (often called gaps).

In this letter we consider identical with [3] physical situation but for more detailed description in NSE introduced the saturation parameter [11].

## Model

We consider the propagation of laser beam at the interface between uniform media and periodic media. In this case we can use the nonlinear Schrödinger equation with dimensionless complex amplitude of the light field $q$:

$$iq_y + \frac{1}{2}q_{xx} + \frac{q|q|^2}{1+S|q|^2} = -R(x)q. \qquad (1)$$

Here $x$ is transverse, and $y$ is longitudinal coordinates which are scaled in terms of beam width and diffraction length; $S$ is the saturation parameter. Function

$$R(x) = -\frac{p}{4}(1-\cos(\Omega x)) \qquad (2)$$

consider total refractive index profile; $p$ is the depth of periodic part of the lattice, $\Omega$ is its frequency.

The saturation parameter are characterized the smoothing of occupancy of two energy levels of atoms under radiation. The saturation effect leads to the dependence of refractive index from the intensity of radiation.

We seek the solution of the equation (1) in such form

$$q(x,y) = f(x)\cdot\exp(i(kx+\omega y)). \qquad (3)$$

Here $f(x)$ is complex periodic function (with period $2\pi/\Omega$), $k$ is the Bloch wave number, $\omega$ is real propagation constant. Then from (1)

$$f_{xx} + 2ikf_x + (k^2 + 2\omega)f + \frac{2f|f|^2}{1+S|f|^2} = -2R(x)f. \qquad (4)$$

Hamiltonian of (4)

$$H = \frac{1}{2}\int_{-\infty}^{+\infty} d\eta \cdot \left( \left|\frac{df}{dx}\right|^2 - 2R(x)|f|^2 - \frac{1}{S^2}\ln\left(1+S|f|^2\right) + \frac{1}{S}|f|^2 \right). \tag{5}$$

Search localized soliton solutions in the following form:
$$f(x) = \eta(x) + i\xi(x),$$
were $\eta(x)$ and $\xi(x)$ represent real and imaginary parts of function $f(x)$ respectively. Then we find from (4) next system of equations:

$$\begin{cases} \eta_{xx} - 2k\xi_x + (k^2 + 2\omega)\eta + \dfrac{2(\eta^2 + \xi^2)\eta}{1 + S(\eta^2 + \xi^2)} = -2R(x)\eta, \\ \xi_{xx} + 2k\eta_x + (k^2 + 2\omega)\xi + \dfrac{2(\eta^2 + \xi^2)\xi}{1 + S(\eta^2 + \xi^2)} = -2R(x)\xi. \end{cases} \tag{6}$$

This system was solved numerically.

Physical solutions correspond to the case
$$k^2 - \omega > 0 \tag{7}$$
(ref. [12])

It is known that periodical wave with weak nonlinearity unstable relative to small modulation perturbations. This nonlinearity called Benjamin-Feir [13].

## Results and discussion

In [3] the linearized version of (4) was solved. Here authors found the domain of existence of gap solitons. It is $\omega > 0$. Also in [3] authors found Floquet-Bloch spectrum.

The results of numerical calculation of (5) are presented below. In case of small S (S=0.05) and in domain $\eta > 0$ two peaks of amplitude and wave function are appeared. In domain $\eta < 0$ waves with the same amplitude is existence only. Profiles of imaginary and real part of solution are shifted on ≈1/8Ω. Figure 2 shows results obtained in case p=2.5. See that two peaks of amplitude and wave function appear also. In case of S=5 we receive identical with fig.1-2 results.

From figures 1-2 see that in uniform media solutions of (5) are oscillated waves. Presence of periodic media is changed the solutions. But surface gap solitons are spatially localized despite the defocusing character of the nonlinearity of both medias.

Gap surface solitons are formed when forward and backward propagating waves experience Bragg scattering in the lattice.

Figure 3 are shown the profile of wave function and the profile of refractive index which was modulated by function R(x). See that two peaks in wave function profile are appeared between two maximums of function R(x). Also amplitude of the first function is smaller than the amplitude of second function.

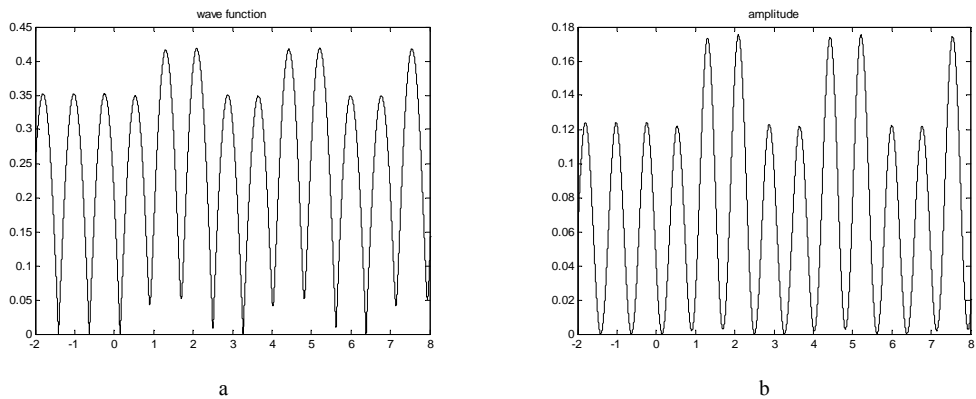

a                                      b

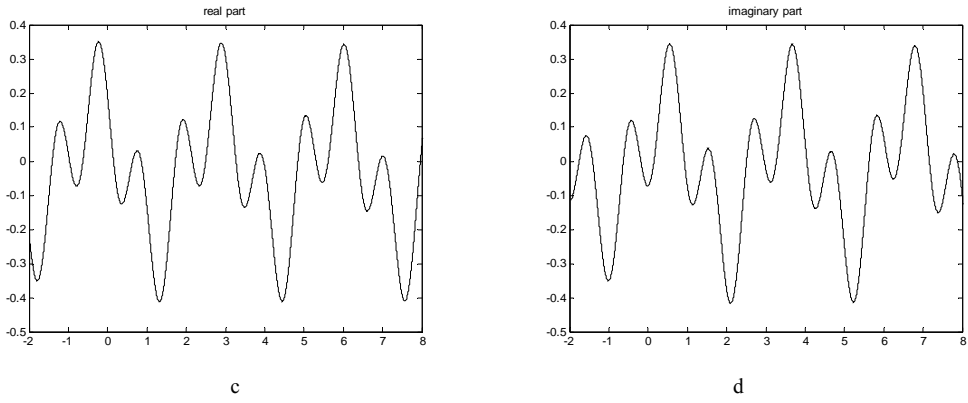

c | d

Figure 1. a. wave function, b. amplitude, c. real part, d. imaginary part
p=0.5, Ω=4, S=0.05

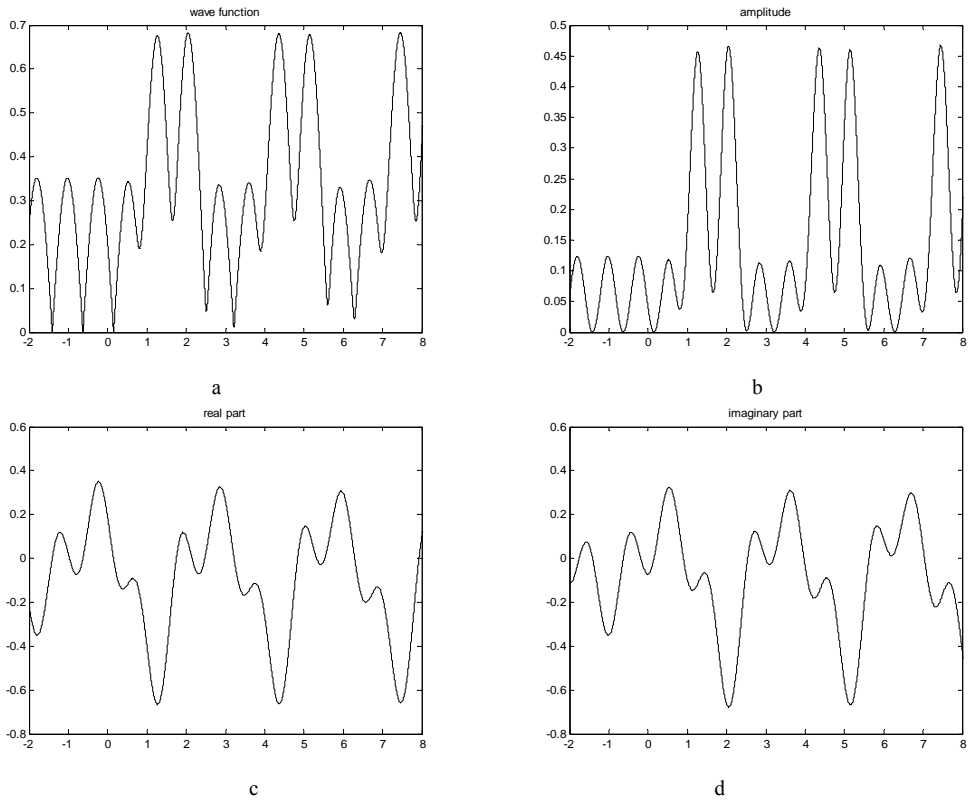

a | b

c | d

Figure 2. a. wave function, b. amplitude, c. real part, d. imaginary part
p=2.5, Ω=4, S=0.05

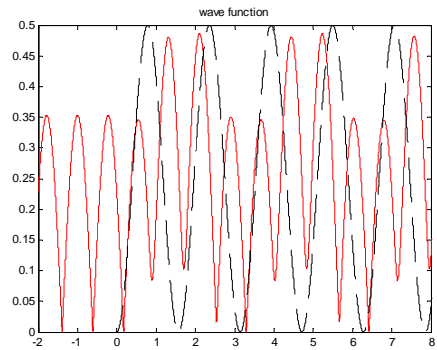

Figure 3. Lattice potential and wave function profile

*Institute of Physics National Ukrainian Academy of Sciences*
unitedlevko@yandex.ru